\documentclass[runningheads]{llncs}
\usepackage{graphicx}
\usepackage{textcomp}
\usepackage{comment}
\usepackage{todonotes}
\usepackage{color}
\usepackage{setspace}
\usepackage{amsmath}

\begin{document}
\title{Uncertainty-Aware Training for Cardiac Resynchronisation Therapy Response Prediction}

\titlerunning{Uncertainty-Aware Training for CRT Response Prediction}

\author{Tareen Dawood\inst{*1}\and
Chen Chen\inst{3} \and
Robin Andlauer \inst{1} \and
Baldeep S. Sidhu \inst{1,2} \and
Bram Ruijsink \inst{1,2} \and
Justin Gould \inst{1,2} \and
Bradley Porter \inst{1,2} \and
Mark Elliott \inst{1,2} \and
Vishal Mehta \inst{1,2} \and
C. Aldo Rinaldi \inst{1,2} \and
Esther Puyol-Ant\'on\inst{1} \and
Reza Razavi\inst{1,2} \and 
Andrew P. King\inst{1}}

\authorrunning{T.Dawood et al.}

\institute{School of Biomedical Engineering \& Imaging Sciences, King's College London, UK\and 
Guys and St Thomas' Hospital, London, UK \and
BioMedIA Group, Department of Computing, Imperial College London, UK}

\maketitle 
\vspace{-2mm}
\begin{abstract}
Evaluation of predictive deep learning (DL) models beyond conventional performance metrics has become increasingly important for applications in sensitive environments like healthcare. Such models might have the capability to encode and analyse large sets of data but they often lack comprehensive interpretability methods, preventing clinical trust in predictive outcomes. Quantifying uncertainty of a prediction is one way to provide such interpretability and promote trust. However, relatively little attention has been paid to how to include such requirements into the training of the model.
In this paper we: (i) quantify the data (aleatoric) and model (epistemic) uncertainty of a DL model for Cardiac Resynchronisation Therapy response prediction from cardiac magnetic resonance images, and (ii) propose and perform a preliminary investigation of  an uncertainty-aware loss function that can be used to retrain an existing DL image-based classification model to encourage confidence in correct predictions and reduce confidence in incorrect predictions. Our initial results are promising, showing a significant increase in the (epistemic) confidence of true positive predictions, with some evidence of a reduction in false negative confidence.
\keywords{Uncertainty  \and Cardiac Resynchronisation Therapy \and Reliability \and Trust \and Awareness.}
\end{abstract}

\section{Introduction}
Cardiac Resynchronisation Therapy (CRT) is an established treatment for advanced heart failure (HF) patients \cite{Katbeh2019CardiacRT}. Current clinical guidelines for CRT patient selection use a limited set of characteristics: New York Heart Association classification, left ventricular ejection fraction, QRS duration, the type of bundle branch block, etiology of cardiomyopathy and atrial rhythm \cite{authors20132013}. However, despite this a significant number of patients (approximately $1$ in $3$) gain little or no symptomatic benefit from CRT \cite{Katbeh2019CardiacRT}. The variability in CRT response has gathered clinical interest, with recent research identifying several factors that are associated with CRT response. For example, clinical studies have shown that apical rocking, septal flash and myocardial scarring are all associated with CRT response \cite{stuatescu2021cardiac}.

In parallel, other researchers have sought to leverage the power of machine learning (ML) techniques in predicting response to CRT. Peressutti \emph{et al} \cite{peressutti2017framework} used supervised multiple kernel learning (MKL) to develop a predictive CRT response model incorporating both motion (obtained from cardiac magnetic resonance (CMR) images) and non-motion data.
Cikes \emph{et al} \cite{cikes2019machine} showed that unsupervised MKL and multimodal data (clinical parameters and image data) can be used to phenotype HF patients and identify which groups benefit the most from CRT. Puyol-Ant\'{o}n \emph{et al} \cite{puyol2020interpretable} developed a deep learning (DL)-based model based on CMR imaging that predicted CRT response and also provided explanatory concepts to aid interpretability.

In recent years, DL models have dominated medical research, but they are often developed without consideration of how the models will be used in clinical practice. In most cases where a DL model could potentially be used for complex predictive problems (such as CRT patient selection), it is unlikely that the model will be used as a stand-alone ``black box'' tool to replace clinicians. Rather, it will act as decision-support tool to aid clinical decision making. This consideration raises the important issue of clinical trust in the model \cite{elton2020self}. 

One way to develop clinical trust in predictive outcomes is to quantify the confidence of a model in its automated decision. Two identified sources of uncertainty are aleatoric uncertainty, which is caused by noisy data inputs and epistemic uncertainty, which refers to uncertainties inherent in the model itself \cite{hullermeier2021aleatoric}. Abdar \emph{et al} \cite{abdar2021review} performed a comprehensive review of the  techniques, applications and challenges of uncertainty quantification in DL. Interestingly, the literature indicates that uncertainty quantification has been predominantly applied to segmentation applications and less so for disease diagnosis or treatment response predictions. We believe that uncertainty information is an important measure to include when providing cardiologists with an automated decision-support tool, in order for them to start developing trust and effectively utilising the tool to aid their daily clinical work.

To date, whereas a number of works have quantified uncertainty in DL models, relatively little attention has been paid to including knowledge of uncertainty in the training of such models. For example, in CRT response prediction, confident incorrect predictions should be avoided, whereas less confident incorrect predictions are less of a problem. Likewise, for correct predictions it is preferred that they should be confident rather than less confident. It would be desirable for a predictive model to take these requirements into account, rather than purely focusing on prediction accuracy.

Related work to this idea includes Geifman and El-Yaniv \cite{geifman2017} who proposed a method for `selective' image classification, in which only confident cases are attempted based on a user-defined risk level. Recent work by Ding \emph{et al} proposed uncertainty-aware training for segmentation tasks by incorporating an additional loss term that aimed to maximise performance on confident outputs \cite{ding2020uncertainty}. In this paper we draw inspiration from Ding \emph{et al} and propose a novel uncertainty-aware training loss for a classification problem, and investigate its effect on the aleatoric and epistemic uncertainties of a DL-based CRT response prediction model.

Our key contributions in this paper are: (1) for the first time, we quantify the epistemic and aleatoric uncertainties of a state-of-the-art DL-based CRT response prediction model, and (2) to the best of our knowledge this is the first investigation of an uncertainty-aware training method for a classification problem in medical imaging.

\section{Materials}

We used two databases to train and evaluate our uncertainty-aware CRT prediction model. First, the CMR short-axis (SA) stacks of 10,000 subjects (a mix of healthy and cardiovascular patients) from the UK Biobank dataset \cite{petersen2015uk} were used for pre-training of the segmentation and variational autoencoder (VAE) models described in Section \ref{sect:methods}. A second database from the clinical imaging system of Guy’s and St Thomas’ NHS Foundation Trust (GSTFT) Hospital was also used, consisting of 20 heart failure (HF) patients and 73 CRT patients. All 73 CRT patients met the conventional criteria for CRT patient selection. The HF patients were used to fine tune the VAE. The 73 CRT patients were used to train and evaluate the baseline and uncertainty-aware CRT prediction models (i.e. VAE and classifier, see Section \ref{sect:methods}.)

For the CRT patients, CMR imaging was performed prior to CRT and the CMR multi-slice SA stack was used in this study. The Siemens Aera 1.5T, Siemens Biograph mMR 3T, Philips 1.5T Ingenia and Philips 1.5T and 3T Achieva scanners were used to perform CMR imaging. The typical slice thickness was 8-10mm, in-plane resolution was between $0.94 \times 0.94 \mbox{mm}^2$ and $1.5 \times 1.5 \mbox{mm}^2$ and the temporal resolution was 13–31 ms/frame. Using post-CRT echocardiography images (at 6 month follow up), a positive response was defined as a 15\% reduction in left ventricular (LV) end-systolic volume.

Institutional ethics approval was obtained for use of the clinical data and all patients consented to the research and for the use of their data. All relevant protocols were adhered to in order to retrieve and store the patient data and related images. 
For all CMR datasets, the top three slices of the SA stack were employed as the input to the model described in Section \ref{sect:methods}.
All slices were spatially re-sampled to $80 \times 80$ pixels and temporally re-sampled to $T=25$ time samples before being used for training/evaluation of the models.

\section{Methods}
\label{sect:methods}

As a baseline CRT response prediction model we used the state-of-the-art DL-based approach proposed by Puyol-Ant\'{o}n \emph{et al} \cite{puyol2020interpretable}. This consists of a segmentation model to automatically segment the LV blood pool, LV myocardium and right ventricle (RV) blood pool from the 3 CMR SA slices at all time frames over the cardiac cycle. These segmentations are then used as input to a VAE with a classification network to predict CRT response from a concatenation of the latent representations for each time frame. Details of training the segmentation and VAE/classifier models are provided in Section \ref{sect:imptooleval}. Note that we do not include the explanatory concept classifier proposed in \cite{puyol2020interpretable} in our baseline as we wish to concentrate our analysis on CRT response prediction confidence in this paper (i.e. we use the baseline model only from \cite{puyol2020interpretable}). The architecture of this model is illustrated in Figure \ref{fig:model}.

\subsection{Uncertainty-aware Loss Function}

The baseline CRT prediction model utilised a loss function that combined the standard reconstruction and Kullback-Leibler terms for the VAE and a term for the CRT response classifier. In this work we propose an additional term for uncertainty-aware training, inspired by \cite{ding2020uncertainty}. The final loss function used was:

\begin{equation}
    \mathcal{L}_{\mathrm{total}}
    = \mathcal{L}_{\mathrm{re}} 
    + \beta \mathcal{L}_{\mathrm{KL}}
    + \gamma \mathcal{L}_{\mathrm{C}}
    + \alpha \mathcal{L}_{\mathrm{U}}
    \label{eqn:all_loss}
    \end{equation}

where $\mathcal{L}_{\mathrm{re}}$ is the cross-entropy between the input segmentations to the VAE and the output reconstructions, $\mathcal{L}_{\mathrm{KL}}$ is the Kullback-Leibler divergence between the latent variables and a unit Gaussian,
$\mathcal{L}_{\mathrm{C}}$ is the binary cross entropy loss for the CRT response classification task and 
$\mathcal{L}_{\mathrm{U}}$ is the uncertainty-aware loss function. 
$\beta$, $\gamma$ and $\alpha$ are used to weight the level of influence each term has to the total loss. 

The uncertainty-aware loss term is defined as:
\begin{eqnarray}\centering
    \mathcal{L}_{\mathrm{U}}
    = & \frac{1}{N_{FP}} \sum_{i \in \text{FP}, j \in \text{TP}}
    \max({\mathcal{P}_{\mathrm{i}}^{+ve} - \mathcal{P}_{{j}}^{+ve} + m,0}) \nonumber \\
    & + \frac{1}{N_{FN}} \sum_{i \in \text{FN}, j \in \text{TN}}
    \max({\mathcal{P}_{\mathrm{i}}^{-ve} - \mathcal{P}_{{j}}^{-ve} + m,0})
    \label{eqn:uncertaintyloss}
\end{eqnarray}

Here, $\mathcal{FP}$, $\mathcal{TP}$, $\mathcal{FN}$ and $\mathcal{TN}$ represent sets of samples from a training batch which are classified as false positives, true positives, false negatives and true negatives respectively, and $N_{FP}$ and $N_{FN}$ are counts of the number of false positives and false negatives in the batch. $\mathcal{P}^{+ve}$ and $\mathcal{P}^{-ve}$ represent the class probabilities of the classifier (after the Softmax layer) for positive and negative response, i.e. $\mathcal{P}^{+ve} = 1 - \mathcal{P}^{-ve}$. The subscripts for these terms represent the sample used as input. Intuitively, Eq. (\ref{eqn:uncertaintyloss}) evaluates all pairs of correct/incorrect positive/negative predictions in a training batch, and the terms will be positive when the incorrect prediction ($i$) has a higher confidence than the correct one ($j$). If the correct one has a higher confidence than the incorrect one by a margin of the hyperparameter $m$ or more it will be zero.

\subsection{Uncertainty Quantification}
\label{sect:uncertquant}

\begin{figure}[ht]%
    \centering
    \includegraphics[width=\textwidth]{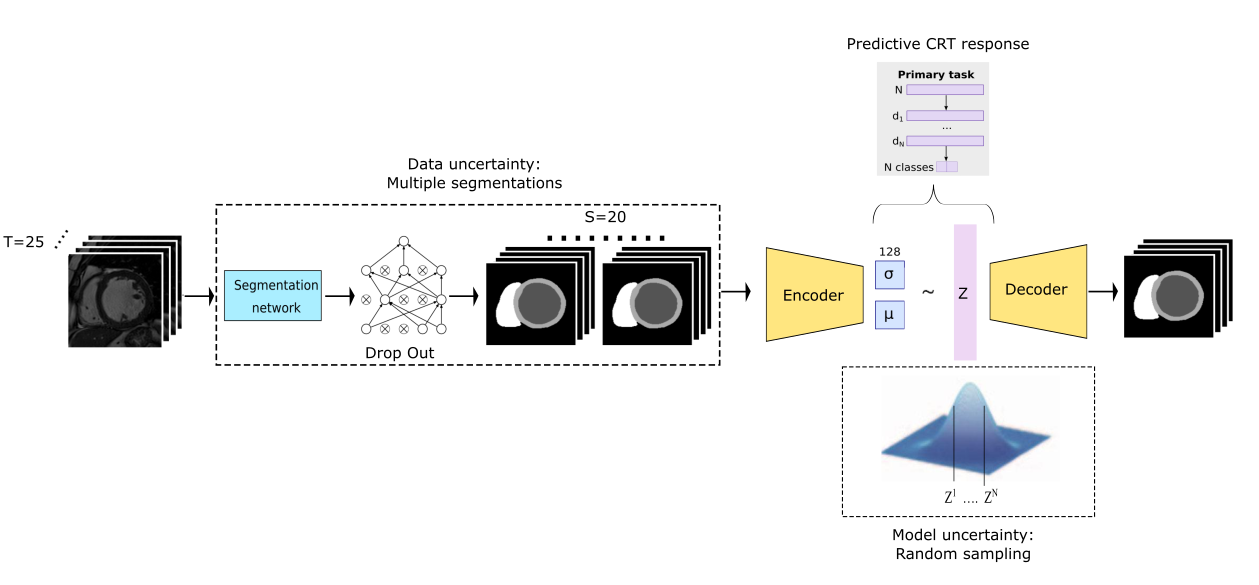}
    \caption{Diagram showing the architecture of the baseline VAE/classification model developed by Puyol-Ant\'{o}n \emph{et al} \cite{puyol2020interpretable} and the stages at which uncertainty estimates were made. The CMR SA images are segmented using a DL-based model, and these segmentations act as the inputs to the VAE. The classification is performed in the latent space of the VAE and points at which aleatoric uncertainty and epistemic uncertainty were estimated are shown in the dotted blocks.}
    \label{fig:model}%

\end{figure}

To evaluate the uncertainty characteristics of the baseline CRT prediction model and assess the impact of our uncertainty-aware loss function we estimate the aleatoric and epistemic uncertainties. The specific points at which uncertainty was estimated are illustrated in Figure \ref{fig:model}. To estimate the aleatoric uncertainty of the CRT response prediction model we generated multiple plausible segmentation inputs to the VAE using inference-time drop out in the segmentation model with probability=$0.5$ in the decoder layers.
Aleatoric uncertainty was then estimated using the CRT response prediction of the original data's segmentations and those from  19 additional segmentation sets generated in this way, i.e. the original and 19 additional segmentations were propagated through the VAE and CRT classifier.

The epistemic uncertainty of the CRT response prediction model was estimated using random sampling in the latent space of the VAE. Again, the original embedding together with 19 additional random samples were used for estimating epistemic uncertainty. Increasing the number of samples from the latent space did not have a statistically significant difference on the estimate but did adversely affect simulation time, therefore just 20 samples were used for epistemic uncertainty estimation. For both types of uncertainty, the CRT response predictions were made for the 20 samples and used to compute a response prediction confidence as the percentage of positive responses out of the 20.

\vspace{-2mm}

\subsection{Tools, Implementation and Evaluation}
\label{sect:imptooleval}

We implemented all methods using the Python programming language, the PyCharm IDE and the open source PyTorch tensor library. All models were trained on a NVIDIA A6000 48GB using an Adam optimiser with a faster learning rate for the VAE and a slower rate for the CRT classifier ($\mathrm{10^{-2}}$) to ($\mathrm{10^{-8}}$), with a batch size of 8. 

The segmentation model was pre-trained using the 10,000 UK Biobank subjects \cite{petersen2015uk}. We then fine-tuned the model using 300 slices (multiple slices/time points from 20 CMR scans) from the HF patients. The VAE was pre-trained using the UK Biobank data and fined tuned with the HF data and then both the VAE and CRT classifier were trained with the CRT patient data for 300 epochs. For training the uncertainty-aware model, the fine-tuned VAE model was used, the uncertainty-aware loss function introduced and then both the VAE and CRT classifier trained for 300 epochs using the 73 CRT patients.

Both the baseline model and the uncertainty-aware model were trained and evaluated using a 5-fold nested cross validation. For each of the 5 outer folds, an inner 2-fold cross validation was performed to optimise hyperparameters using the training data. In these inner folds, the set of hyperparameters yielding the highest accuracy was selected. The optimal hyperparameters were used to train a model (using all training data for that fold) which was then applied to the held-out (outer) fold. This process was repeated for all outer folds. In this way, hyperparameter optimisation was performed using training data and the model was always applied to completely unseen data. Note also that the CRT data had not been used in pre-training either the segmentation model or the VAE. The hyperparameters optimised in this way were (see Eq. (\ref{eqn:all_loss})):  $\beta$ ($0.001-2$), $\gamma$($0-2$), $\alpha$ ($0.01-2$), $m$ ($0.01-1$) and the size of hidden layers in the CRT classifier ($32,64$). The final optimal parameters for the baseline model were, $\beta$ ($0.1$), $\gamma$ ($0.6$) and for the uncertainty-aware model, $\beta$ ($0.001$), $\gamma$($0.5$), $\alpha$ ($0.05$) and $m$ ($0.6$). Both models were optimal with 32 hidden layers in the CRT classifier. 

\section{Experiments and Results}

To evaluate the impact of our uncertainty-aware loss function, we first measured the overall classification accuracy of the baseline and then the uncertainty-aware model and achieved a balanced accuracy (i.e. average of sensitivity and specificity) of 70\% and 68\% respectively. We note that this was lower than the value reported in \cite{puyol2020interpretable}, as a result of implementing nested cross validation, which prevents leakage of data between the training and testing sets.

We further summarise the epistemic and aleatoric uncertainties of the outputs (estimated as described in Section \ref{sect:uncertquant}). The results are shown in Tables \ref{table:1} and \ref{table:2} for epistemic and aleatoric uncertainty respectively. These tables both show counts of the numbers of true positives (TP), false negatives (FN), false positives (FP) and true negatives (TN) in 3 confidence bands. Note that these bands represent confidence in the predicted response, i.e. for the subjects predicted as responders it is the percentage confidence in positive response, and for subjects predicted as non-responders it is the percentage confidence in negative response (which is 100 minus the positive response confidence).
\vspace{10mm}

\begin{table}[htb]\centering
\caption{Summary of epistemic (model) uncertainty for the baseline and uncertainty-aware models. (GT=Ground Truth, TP=True positive, FN=False Negative, FP=False positive, TN=True negative)}
 \begin{tabular}{||c || c | c ||c |c |c | c||}
 \hline
 \multicolumn{5}{|c|} {Baseline model} \\
 \hline
 Confidence& \multicolumn{2}{|c||}{GT +ve} &  \multicolumn{2}{|c|}{GT -ve}\\
 \hline
 & Pred. +ve (TP) & Pred. -ve (FN) & Pred. +ve (FP) & Pred. -ve (TN) \\
 \hline
 0-30 & 1 & 3 & 10 & 6 \\ 
 31-70 & 11 & 5 & 2 & 7 \\
 71-100 & 23 & 2 & 0 & 3 \\ [0ex] 
 \hline
 \multicolumn{5}{|c|}{Uncertainty-aware model} \\
 \hline
 0-30 & 3 & 5 & 14 & 4 \\ 
 31-70 & 6 & 1 & 0 & 7 \\
 71-100 & 27 & 3 & 0 & 3 \\ [0ex] 
 \hline
 \end{tabular}
 \label{table:1}
\end{table}
\vspace{5mm}

\begin{table}[!htb]\centering
\caption{Summary of aleatoric (data) uncertainty for the baseline and uncertainty-aware models. (GT=Ground Truth, TP=True positive, FN=False Negative, FP=False positive, TN=True negative) }
 \begin{tabular}{||c || c | c ||c |c |c | c||}
 \hline
 \multicolumn{5}{|c|}{Baseline model} \\
 \hline
 Confidence& \multicolumn{2}{|c||}{GT +ve} &  \multicolumn{2}{|c|}{GT -ve}\\
 \hline
 & Pred. +ve (TP) & Pred. -ve (FN) & Pred. +ve (FP) & Pred. -ve (TN) \\
 \hline
 0-30 & 3 & 4 & 11 & 6 \\ 
 31-70 & 6 & 0 & 1 & 4 \\
 71-100 & 26 & 6 & 0 & 6 \\ 
 \hline
 \multicolumn{5}{|c|}{Uncertainty-awareness trained model} \\
 \hline
 0-30 & 3 & 5 & 13 & 7 \\ 
 31-70 & 3 & 0 & 0 & 4 \\
 71-100 & 30 & 4 & 1 & 3 \\ [0ex] 
 \hline
 \end{tabular}
 \label{table:2}
\end{table}

The results show that, for epistemic and aleatoric uncertainty, the uncertainty-aware model encouraged confidence in correct predictions (i.e. see the true positives column). There is also some evidence of reduced confidence in incorrect predictions for the non-responders for epistemic uncertainty. However, overall significant changes were not as evident for false negatives for both aleatoric and epistemic uncertainty. Further analysis is required to develop a more robust method to improve the outcomes, however, as a preliminary investigation the results are encouraging and suggest the need to include uncertainty awareness training when building prediction models.
\vspace{12mm}

\section{Discussion and Conclusion}

In this paper we have proposed a novel uncertainty-aware loss term for classification problems, in which we envisage the model being used as a decision-support tool for clinicians. In such situations, to promote clinical trust and to maximise the utility of the tool, it is imperative that confident incorrect predictions are minimised. To illustrate the effects of including the developed uncertainty-aware loss function, aleatoric and epistemic uncertainty were estimated for a baseline CRT response prediction model and the model that utilised the new loss term. 

The results indicate an increase in confident true positive predictions for both epistemic and aleatoric uncertainty.
There is also some evidence of a reduction in confidence of incorrect predictions, particularly in epistemic uncertainty. However, despite these promising signs the method requires further investigation and refinement. One weakness in the proposed framework is that our uncertainty-aware loss term is based on the class probabilities of the classifier network which are estimates of epistemic uncertainty and it is known that these probabilities can overestimate confidence in predictions \cite{hendrycks2016baseline}. It is not possible to directly include uncertainties estimated using sampling based approaches such as inference-time drop out due to them not being differentiable. In future work we will address this issue by seeking to incorporate more reliable and differentiable estimates of uncertainty, such as the use of soft labels, to reduce overestimation of confidence. Further investigation into uncertainty estimation approaches during training of DL models should be explored to develop a more robust and accurate method for uncertainty-aware training.

Future work will also investigate methods to improve the aleatoric uncertainty outcomes, such as the implementation of a probabilistic U-net segmentation model \cite{kohl2018probabilistic} instead of using the inference-time drop out technique, which may produce more realistic segmentations and improved estimates of aleatoric uncertainty. Additionally, by incorporating electrical and mechanical parameters with the existing model similar to \cite{albatat2018electromechanical} the CRT prediction accuracy could be improved. In their paper, Albatat \emph{et al} \cite{albatat2018electromechanical} used mesh generation to develop an electromechanical model that could be used for patient specific CRT prediction. This approach alongside investigating alternate classification model architectures, and potentially expanding the size of the data set, may improve the performance achieved with the developed uncertainty-aware loss function, to improve the accuracy and reliability of CRT predictions. 

We believe that the preliminary investigation and framework we have proposed represents an important step on the road towards clinical translation of  DL models for complex predictive tasks, such as CRT patient selection. Although we demonstrated the framework in this paper for CRT response prediction we believe it will be applicable to a range of other complex diagnostic/prognostic tasks for which DL models will likely be used for decision-support. In future work, we aim to evaluate the impact of our confidence estimates on clinical decision making with more robust methodologies. 

\section {Acknowledgements}
This work was supported by the NIHR Guys and St Thomas Biomedical Research Centre and the Kings DRIVE Health CDT for Data-Driven Health. This research has been conducted using the UK Biobank Resource under Application Number 17806.The work was also supported by the EPSRC through the SmartHeart Programme Grant(EP/P001009/1).


\bibliographystyle{unsrt}
\bibliography{refs}
\end{document}